\documentclass[number,sort&compress,3p]{elsarticle}

\usepackage{lineno,hyperref}

\usepackage{multirow}
\usepackage{graphicx}
\usepackage{amsmath}
\usepackage{amssymb}
\usepackage{algorithm}
\usepackage{algorithmic}
\usepackage{array}
\usepackage{subfigure}
\usepackage{amsthm}
\theoremstyle{definition}
\newtheorem{definition}{Definition}[section]

\newtheorem{Method}{Method}
\usepackage{epstopdf}
\usepackage{graphicx,times,amsmath}
\usepackage{lastpage,fancyhdr,graphicx}
\modulolinenumbers[5]
\journal{Journal of Computational Information Systems}
\usepackage{numcompress}
\begin{document}
	
	\begin{frontmatter}
		\title{Basketball Player's Value Evaluation by a Networks-based Variant Parameter Hidden Markov Model}
	
		\author[stu]{Xin Du}
		\author[stu,uoft]{Weihong Cai\corref{mycorrespondingauthor}}
		\cortext[mycorrespondingauthor]{Corresponding author}
		\ead{whcai@stu.edu.cn}
		\author[fua,cub]{Jianquan Liu}
		\author[stu]{Ding Yu}
		\author[stu]{Kai Xu}
		\author[stu]{Wei Li}

		\address[stu]{Department of Computer Science, College of Engineering, Shantou University, China}
		\address[uoft]{Big Data Research Institute, Tong Xing Technology Corporation, China}
		\address[fua]{System Platform Research Laboratories, NEC Corporation, Japan}
		\address[cub]{Graduate School of Science and Engineering, Hosei University, Japan}

		\begin{abstract}
Determining the value of basketball players through analyzing the players' behavior is important for the managers of modern basketball teams. However, conventional methods always utilize isolated statistical data, leading to ineffective and inaccurate evaluations. Existing models based on dynamic network theory offer major improvements to the results of such evaluations, but said models remain imprecise because they focus merely on evaluating the values of individual players rather than considering them within their current teams.
To solve this problem, we propose an analysis and evaluation model based on networks and a hidden Markov model. To the best of our knowledge, we are the first to combine a network form representing the players who are playing with the use of a hidden Markov model to mine the network and generate the desired results. Applying our approach to SportVU data collected from the National Basketball Association shows that this analysis and evaluation model can effectively analyze the performance of each player in a game and provides an assistive tool for team managers.
\end{abstract}


		
	\end{frontmatter}
	
	\section{Introduction}\label{sec:introduction}

Behavioral analysis of basketball players and the measurement of their values are crucially important for building an admirable basketball team. Without an elaborate assessment system for basketball players, the team manager is unable to devise a strategy by which to utilize athletes of different types and values~\cite{gudmundsson2017spatio}. Such a system can not only provide suggestions for team managers for formulating strategies but also play important roles in the construction of coaching tactics and in the players' own training.

Conventional systems often use concrete statistical data, such as scoring, assists, and rebounds, to measure player's behavior in a game. On the basis of these data, they study the changes in individual players' physical states in competition and try to divide the players' roles in team sports (such as a leading role versus a supporting role), so as to orient themselves to their players' values~\cite{carling2010analysis,wang2007learning,8634658}. By Svensson's survey report~\cite{svensson2005testing}, such an approach has been proven insufficient insofar as it ignores implicit abilities behind direct data and cannot offer proper strategies by which to improve the total performance of a team with unaltered players. The appearance of the camera on the game court provides opportunities accurately analyzing any specific player in detail. For example, SportVU~\cite{cornacchia2017survey}, which is dedicated to the domain of sports, distinguishes concrete ball passing and records the instantaneous positions of each player. The massive amounts of data produced by similar recording systems enable theoretical analyses with mathematical tools, such as graph theory and hidden Markov modeling, to uncover useful information that can correctly assess players' values.

The players in the games can initially be viewed as isolated points that are connected by ball passing and changing position to form an integrated network, which match the attributes of the network. The first attempt of sports analysis was proposed by Gould and Gatrell~\cite{gould1979structural}, who studied the simplicial complexes of the passing network in the 1977 FA Cup Final. However, their analysis did not receive considerable attention due to the limitations of hardware facilities.

Since the mid-2000s, a series of works have appeared. These works introduced social networks to analyze team sports. Bourbousson et al.~\cite{bourbousson2010team} presented an efficient method for distinguishing the relationships among basketball players on the offensive end using graph theory. Passos et al.~\cite{passos2011networks} provided some insights on graph theory in team sports by analyzing datasets of player interactions during an attacking sequence in water polo. In 2012, Fewell et al.~\cite{fewell2012basketball} introduced a transition graph in basketball games wherein the vertices indicated the five traditional player positions (point guard, shooting guard, small forward, power forward, and center). In this paper, the network properties of degree centrality, clustering, entropy, and flow centrality across teams and positions were used to characterize the game. These results prove that the coordination of players built on local interactions does not necessarily force all players to win the matches. Despite this result, Duch et al.~\cite{duch2010quantifying} still applied the network model to improve the passing accuracy and analysis performance of soccer players. Recently, a series of works by Clemente et al.~\cite{ clemente2015general, clemente2015using} demonstrated that network metrics can be applied to characterize the cooperation of players, which is a powerful tool for helping coaches understand specific properties of their teams and to support decision making to improve the sports training process based on match analysis.

These achievements provide the basis with which to utilize graph theory to evaluate the individual values of players. Indeed, the researchers~\cite{clemente2015using, grund2012network, pena2012network} evaluated the performance of players by network density and heterogeneity. However, it is challenging to evaluate players' exact values on their own team~\cite{albert2000error}. The challenge is that players produce a large amount of data when they interact on the court, and the quantitative analysis of this information cannot be achieved by these scholars' methods~\cite{9284088}. To solve this problem, we use the interactive data between players to construct the network and then apply a machine learning algorithm to analyze these networks.

We introduce a novel criterion to better discover and analyze datasets for evaluating the exact value of a player to his team. This criterion combines three features of players that match the actual situation in a game. The estimation of each player's three features with a machine learning algorithm is the key to analyzing his value to the team. The first feature includes passing efficiency and its specific effect on scoring, which indirectly measures the scores that one player achieves. The second feature involves the probabilities of a person being chosen as an organizer for the next offense. It offers the ability to form an overarching perspective of a player in a game. The last feature is the direct scoring generated by a player, which includes foul shots and three-point shots. Under the novel criterion, we construct a passing network that incorporates the semantics of the game to create an accurate expression of the passing relationship between players to quantify these three features by means of three parameters. The state transition probability indicates the passing relationships, the initial probability represents the probability of a player being chosen as an organizer, and the observation probability illustrates the probabilities of different points that the player scores in a round. To estimate these three parameters, we propose a variant parameter hidden Markov model (VPHMM), which is a parameter estimation paradigm that inherits advantages from hidden Markov models~\cite{Ingle2016Hidden,eddy1998profile,8750782}. 

The main contributions of this paper include the following:
\begin{itemize}
	\item A novel criterion is introduced to evaluate the player's performance from a team-level perspective.
	\item A variant parameter hidden Markov model is proposed to estimate the parameters that indicate three features of the player's performance.
	\item A network-based hidden Markov model is applied to team sports for the first time.
\end{itemize}

The rest of the paper is organized as follows. We describe in detail the methods in Section 2, which includes network analyses and parameter estimation. In Section 3, the results of the measures are empirically demonstrated. Section 4 is the conclusion.
	\section{Methodology}\label{sec:Methodology}
We originally aimed to record and analyze passing networks in basketball to evaluate the exact value of each player to his team. However, because most sources only provide aggregate data across all games, the passing networks were constructed using the parameters trained in our system. When we found that the system, which was created by a hidden Markov model, could train very accurate parameters, we discovered that the evaluation method exceeded our expectations.
All three parameters obtained by the algorithm can play an important role in the network, which is composed of detailed data with which the game players are constructed~\cite{emmert2016fifty}. The subtle performances of a player on the court can be shown by his three parameters. We applied these parameters to basketball games to demonstrate that the approach is feasible and precise in evaluating the value of a basketball player to his team.

\subsection{Network Analyses}

To characterize the basketball matches as a network, the first step is to define the standards by which the players are linked. We construct a finite ${n} \times ${n} network where the vertices represent the players and the edges represent linkages between players (e.g., when a player passes the ball to another player). A ‘round’ can be defined as the moment that the team obtained possession of the basketball to the moment when possession of the basketball was recovered by the opposition~\cite{passos2011networks}. Then, the contribution of a player to a team can be inferred from the individual passing network and, in particular, from centrality and statistical measures, which define the values of a player according to different parameters~\cite{pena2012network}. We provide a novel measure that defines three parameters and discusses their meaning in the context of basketball.

\begin{definition}
(Matrix of state transition probability). Assume that there is a set of players $\mathcal{M} = \left\lbrace 1,2,...,m\right\rbrace$; we define $\omega_{i,j}$ to be the number of passes from player $i$ to player $j$. Thus, the total number of passes in a game should be $\sum_{i=1}^{m} \omega_{i}$. Therefore, the matrix of transition probability $A$ can be solved as follows:
	\begin{equation}
	A_{i,j} = \frac{ \omega_{i,j}}{
		\sum_{i=1}^{m} \omega_{i}},
	\end{equation}
	
	\begin{itemize}
		
		\item For every player $i \in \mathcal{M}$,
		
		\item The player did not pass the ball if $i = j$.
	\end{itemize}

\end{definition}

\begin{definition}
(Matrix of observation probability). Assume that there is a set of observations $\mathcal{N} = \left\lbrace 1,2,...,n\right\rbrace$; we define $b_{j,k}$ to be the number of players $j$ in observation $k$. Thus, the total observations of this player in a game should be $\sum_{j=1}^{n} $b$_{j,k}$. Therefore, the matrix of transition probability $B$ can be solved:
	\begin{equation}
	B_{j,k} = \frac{b_{j,k}}{\sum_{i=1}^{n} b_{i}},
	\end{equation}

	\begin{itemize}
		
		\item For every observation of basketball game $k \in \mathcal{N}$.
		
	\end{itemize}

\end{definition}
\begin{definition}
(Initial probability distribution). We define $\pi$ as the probability distribution of players being chosen as an organizer for the offense.	
\end{definition}

We generated weighted graphs from the probabilities of cumulative passes, which can be obtained from the definition 2.1. Here, previous researchers \cite{fewell2012basketball} usually set $m$ to five letters to represent the five most important players of five basketball positions. Although this better depicts the network diagram by highlighting the important players per the role of the network, it ignores the values of role players. This may distort the truth of the game, and the result will be different from the actual event. The network diagram (see Figure 1) that we set up here does not ignore any one person. The blue circles represent those involved in the match who had playing time. An arrow’s direction represents the pass direction. The origin of the arrow indicates the player who passed the ball, and the arrowhead indicates the player who received the ball. The width and color of the arrows denote the numbers and percentages of passes from one player to another during games. Specifically, thicker arrows represent more passes taking place between players, and thinner arrows illustrate fewer passes occurring among players. The red arrows represent when the number of passes from one player to a specific player is more than 20 percentage of his total passes. When the percentage is between 10 and 20, the arrows are colored orange. Meanwhile, the blue arrows represent when this percentage falls between 5 and 10, and the green arrows denote that ,this percentage is between 1 and 5. The grey arrows represent percentages of less than 1, the minimum percentage.
This weight graph not only takes into account the details but also intuitively reveals the important players in the game. However, when our dataset was analyzed in detail, almost all nodes
were connected, making some nodes difficult to differentiate across
the team's network. To solve this problem, we creatively refined the graph (Figure 2) to each player's perspective. Each player's individual network contains the passes arrows and received arrows. Upon evaluating the values of the players, these individual networks can intuitively indicate the relevance or centrality of a player.

\begin{figure}[!ht] 
	\centering 
	
	\subfigure[]{ 
		\label{fig:subfig_a} 
		\includegraphics[width = 0.52\textwidth]{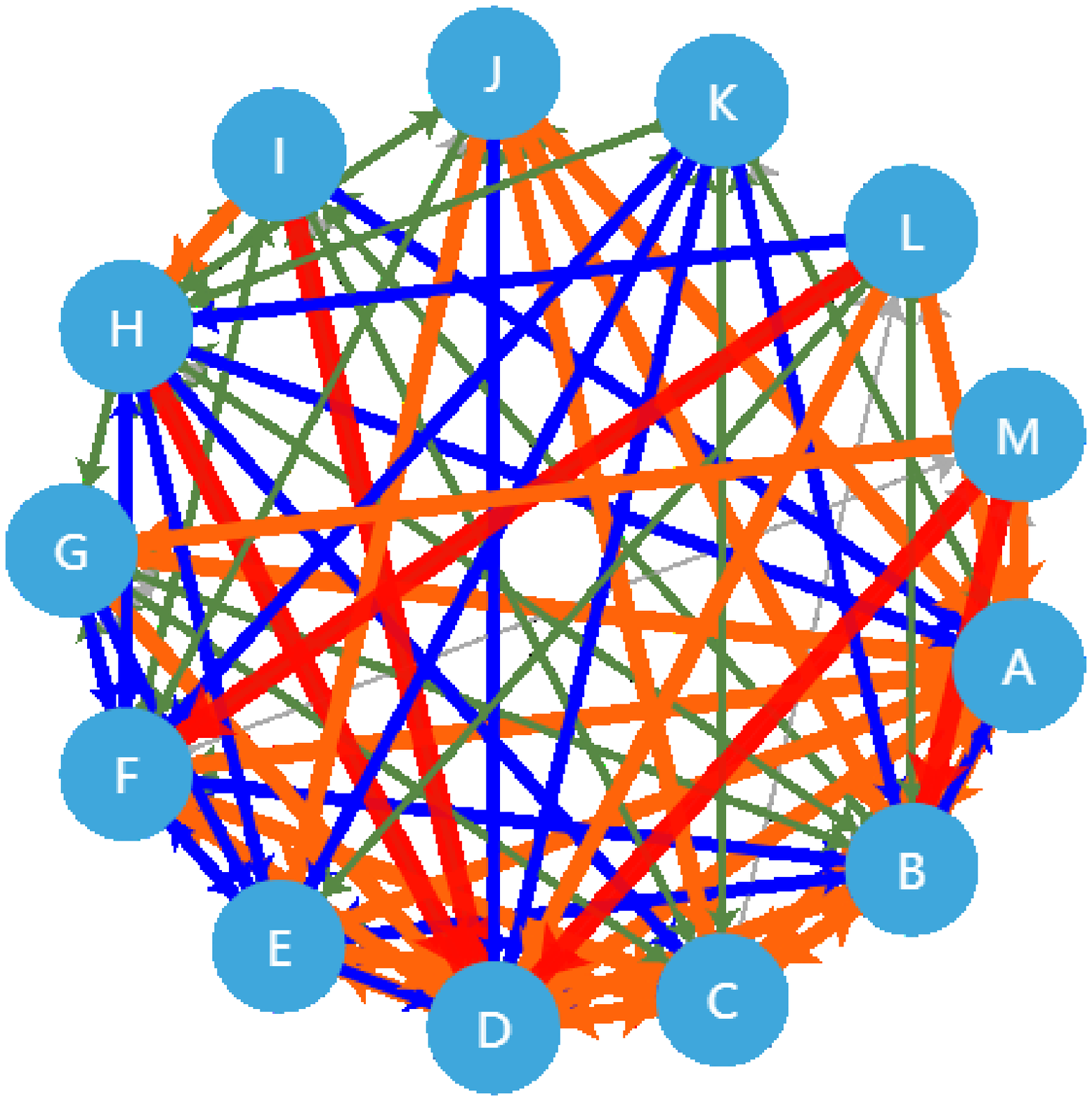} 
	} 
	\hspace{-0.5in}
	\subfigure[]{ 
		\label{fig:subfig_a} 
		\includegraphics[width = 0.52\textwidth]{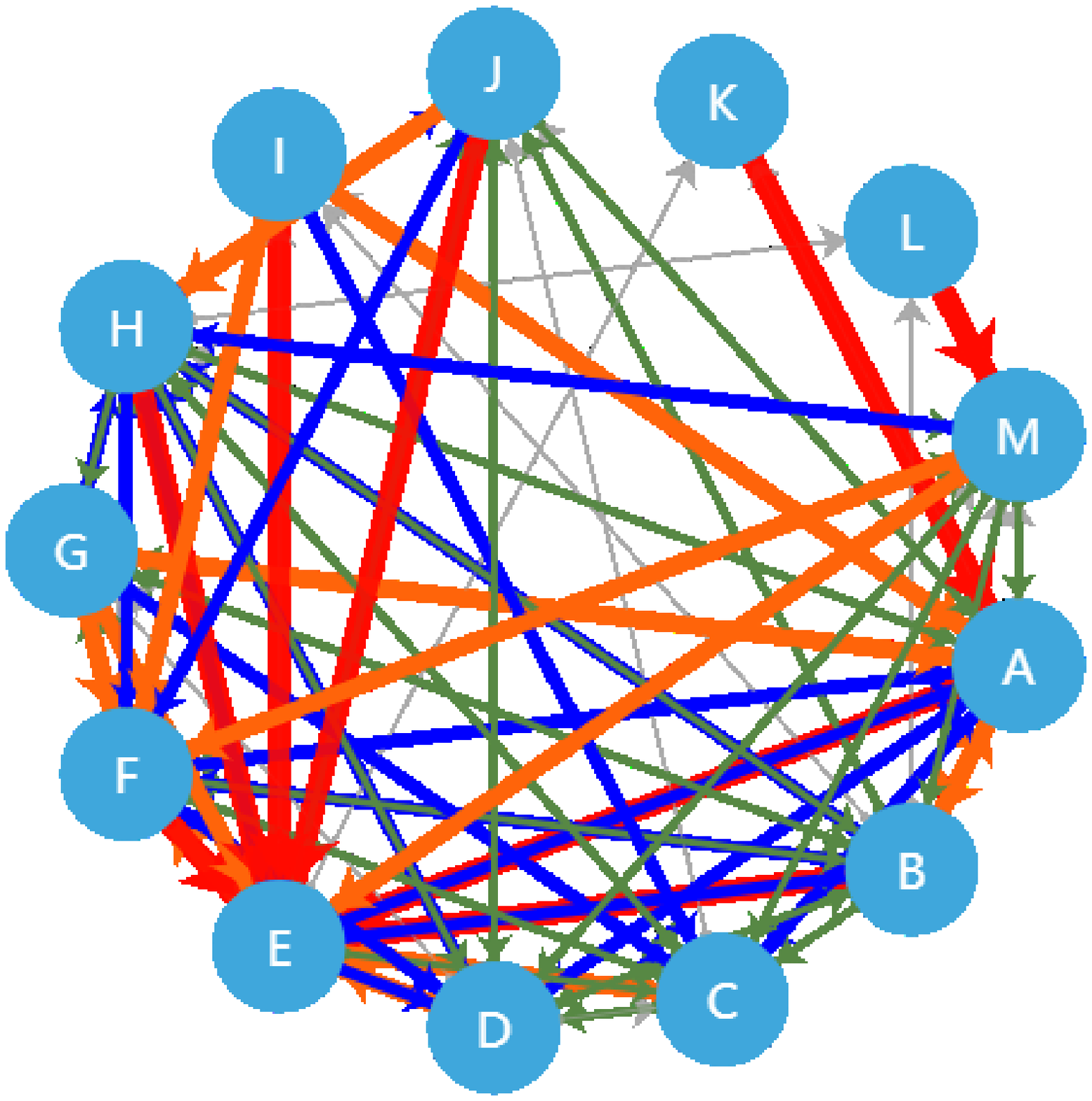} 
	} 
	
	\caption{ {\bf   Network representation for analyzed games played by the Golden State Warriors (a: a win; b: a loss).}
		The blue circles represent the players involved in the game who had playing time. The direction of the arrows represents the pass direction. The origin of the arrow
		indicates the player who passed the ball, and the arrowhead indicates the player who received the ball. The width and color of the arrows denote
		the numbers and percentages of passes from one player to another during games. Specifically, thicker arrows represent more passes having occurred
		between players, and thinner arrows indicate fewer passes having occurred among players. The red arrows indicate that the number of passes from one player
		to the specified player is more than 20 percent of his total passes. When the percentage is between 10 and 20, the arrows are colored orange.
		Likewise, the blue arrows indicate that this percentage is between 5 and 10, and the green arrows indicate that this percentage is between 1 and 5. The gray arrows indicate that the percentage is less than 1, which is the minimum percentage represented by color.} 
	\label{fig1}
\end{figure}

\begin{figure}[!ht] 
	\centering 
	
	\subfigure[]{ 
		\label{fig:subfig_a} 
		\includegraphics[width = 0.51\textwidth]{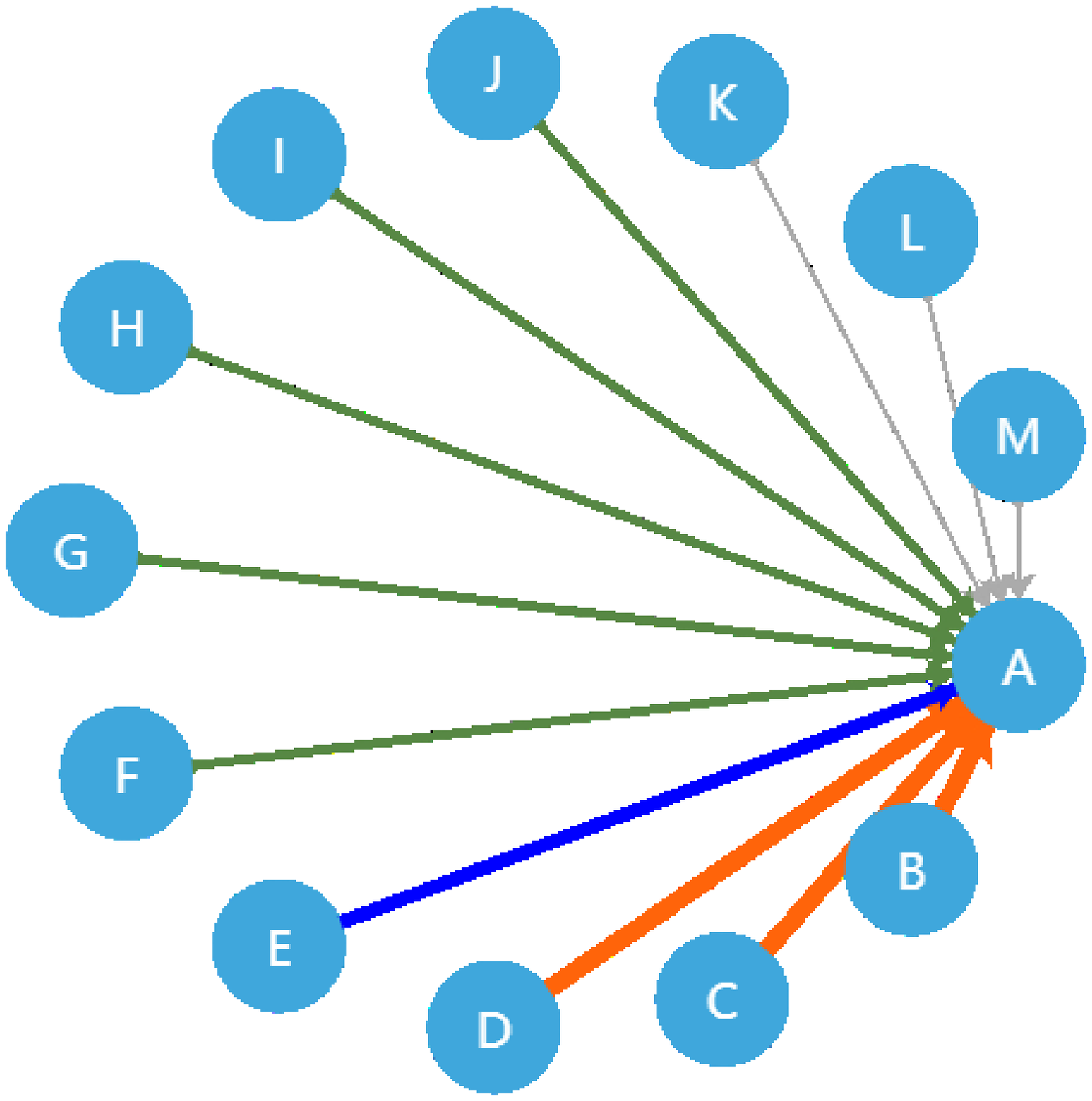} 
	} 
	\hspace{-0.5in}
	\subfigure[]{ 
		\label{fig:subfig_a} 
		\includegraphics[width = 0.51\textwidth]{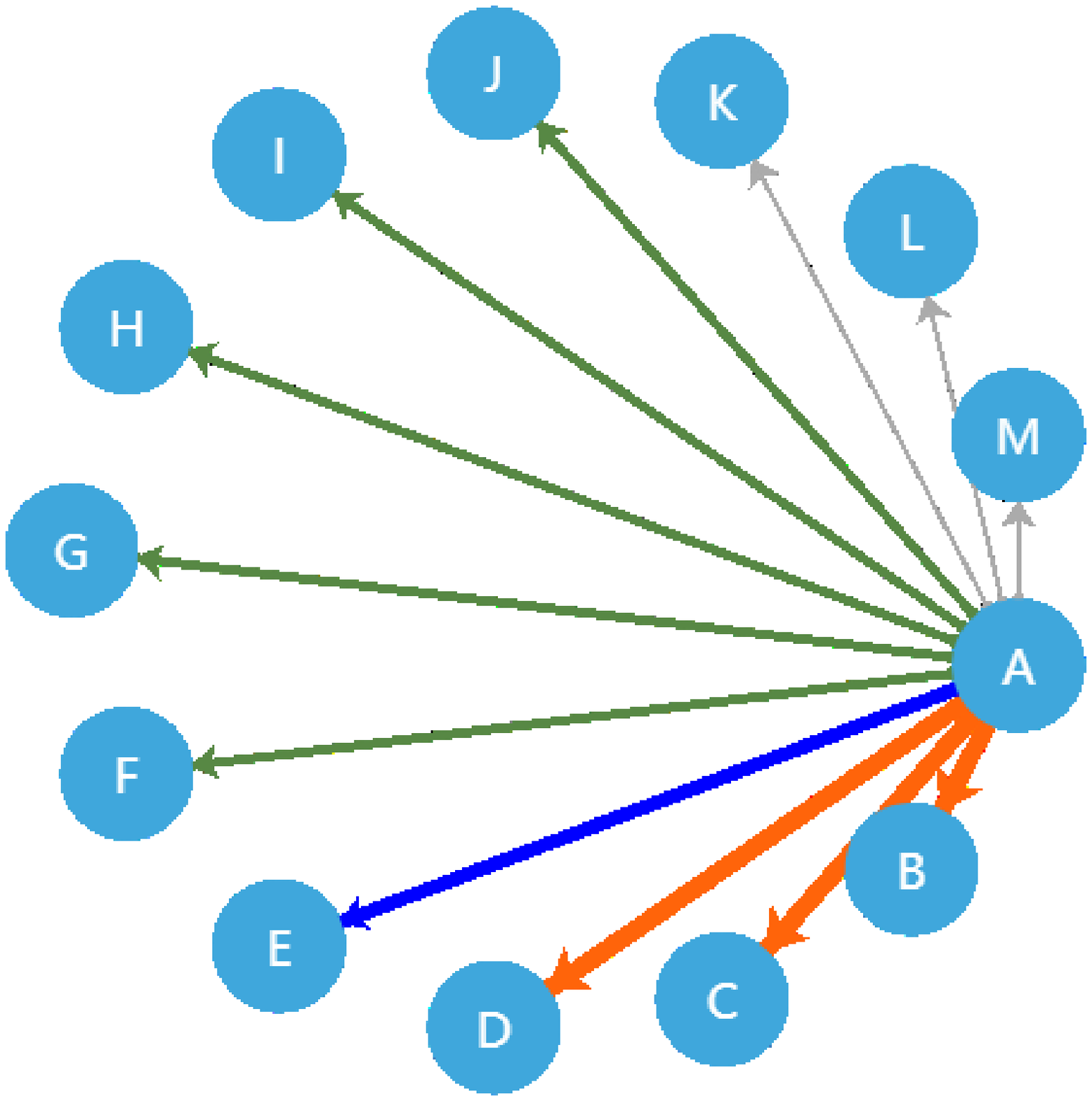} 
	} 
\caption{{\bf Network representation for the analyzed games from the point of view of A (i.e., Stephen Curry).} Fig 2(a) presents the network of A with respect to receiving the ball from others, and Fig 2(b) presents the network of A with respect to passing the ball to other players.} 
	\label{fig2}
\end{figure}

In addition, the method of evaluation also includes the score distribution of the player. The score distribution defines the probability of different numbers of points scored by the player in a round. 
The last parameter we used in our method represents the probability of the player being the organizer. These three parameters trained in our system represent the aforementioned features. We will illustrate the application in detail in Section 3 with an example about basketball.

\subsection{Parameters Training
}

In the definitions of the previous subsection, we defined the following parameters: the matrix of transition probability $A= \{ a_{ij}$\}, the matrix of observation probability $B$ = \{$b_{ij} $\}, and the initial probability distribution $\pi$ = \{$\pi_{i}$\}. We provide the definitions and mathematical procedures of our VPHMM algorithm in this subsection.

We define the players to be the finite set of states M = \{$\alpha_{1}$,$\alpha_{2}$,$\alpha_{3}$,...,$\alpha_{m}$\}, and we define the finite set of observations N =\{$\varGamma_{1}$,$\varGamma_{2}$,$\varGamma_{3}$,...,$\varGamma_{n}$\}. These two parameters, which are easily obtained from the large datasets, are used as input for Algorithm~\ref{alg:A} to estimate the other three parameters (i.e., A, B, and $\pi$). Now, we present a practice example to show how the proposed algorithm works.

\begin{algorithm}
	\caption{Parameters training}
	\label{alg:A}
	\begin{algorithmic}[1]
		\REQUIRE  The finite set of states $M$;   the finite set of observations $N$; The initial value $\lambda^{(0)}=(A^{(0)},B^{(0)},\pi^{(0)})$
		\ENSURE  Matrix of transition probability $A$; Matrix of observation probability $B$; Initial state probability distribution $\pi$; The final value $\lambda^{(n+1)}=(A^{(n+1)},B^{(n+1)},\pi^{(n+1)})$
		\STATE $n = 0$.
		\WHILE{$n < T$}
		\STATE $a_{ij}=\frac{\sum_{t=1}^{T-1}\zeta_t(i,j)}{\sum_{t=1}^{T-1}\gamma_t(i)}$.
		\STATE $b_j(k)^{(n+1)}=\frac{\sum_{t=1,a_1=v_k}^{T}\gamma_t(j)}{\sum_{t=1}^{T}\gamma_t(j)}$.
		\STATE $\pi_i^{(n+1)}=\gamma_1(i)$.
		\ENDWHILE
	\end{algorithmic}
\end{algorithm}

Algorithm~\ref{alg:A} was proposed based on a very classical unsupervised learning algorithm in the hidden Markov model~\cite{leos2017analysis}. We combine the estimation method with the properties of network theory to train the parameters and evaluate the players' values in the basketball matches. The estimation formula, which incorporates the semantics of the basketball game~\cite{wei:2015:pst:2783258.2788598}, is as follows:

In the method, $a_{ij}$ denotes the probability of player $i$ passing to another player $j$. We calculate $a_{ij}$ as follows:
\begin{equation}
a_{ij}=\frac{\sum_{t=1}^{T-1}\zeta_t(i,j)}{\sum_{t=1}^{T-1}\gamma_t(i)}.
\end{equation} 

$b_j(k)^{(n+1)}$ represents the probability of player $j$ obtaining $b$ points in one round. Its estimation formula is 
\begin{equation}
b_j(k)^{(n+1)}=\frac{\sum_{t=1,a_1=v_k}^{T}\gamma_t(j)}{\sum_{t=1}^{T}\gamma_t(j)}.
\end{equation}

$\pi_i^{(n+1)}$ represents the rate of player $i$ being chosen as the organizer, and the parameter can be trained from 
\begin{equation}
\pi_i^{(n+1)}=\gamma_1(i).
\end{equation}

According to~\cite{abdel2014convolutional}, $\gamma_t(i)$ as a conclusion is defined as follows:
\begin{equation}
\gamma_t(i) = P(i_t=q_i|O,\lambda)=\frac{\alpha_t(i)\beta_t(i)}{\sum_{i=1}^N\sum_{j=1}^N\alpha_t(i)a_{ij}b_j(o_{t+1})\beta_{t+1}(j)},
\end{equation}
where $\alpha_t(i)$ and $\beta_t(i)$ are two intermediate variables. Both of these variables converge from one side to another and construct a directed edge to form a recursive structure.

$\zeta_t(i,j)$ is an intermediate node in the $t$ round, which is derived from the forward algorithm~\cite{lecun2015deep}, calculated by the formula with the last node. It can be obtained as follows:

\begin{equation}
\zeta_t(i,j) = \frac{P(i_t=q_i,i_{t+1}=q_j,O|\lambda)}{P(O|\lambda)}=\frac{\alpha_t(i)a_{ij}b_j(o_{t+1})\beta_{t+1}(j)}{\sum_{i=1}^N\sum_{j=1}^N\alpha_t(i)a_{ij}b_j(o_{t+1})\beta_{t+1}(j)}.
\end{equation}

The weighted and directed edge relation , $P(i_t=q_i,i_{t+1}=q_j,O|\lambda)$ between rounds can be calculated as follows:
\begin{equation}
P(i_t=q_i,i_{t+1}=q_j,O|\lambda)=\alpha_t(i)a_{ij}b_j(o_{t+1})\beta_{t+1}(j).
\end{equation}

Then, we present their derivation in detail based on the games of the Golden State Warriors in the 2016 season. In our set, we have the known observation sequence $O = (o_{1},o_{2},o_{3},...,o_{T})$ and $o_{T} \in \mathcal{N}$. 
The states' sequence is $I = (i_{1},i_{2},i_{3},...,i_{T})$ and $i_{T} \in \mathcal{M}$.
In our method, the observation sequence $O$ presents a sequence of rounds in the basketball match, and T represents the number of rounds in a game. The states' sequence is illustrated as the player. In our example,  A is Stephen Curry, and B is another player named Andrew Bogut.
$\lambda$ presents the current estimate (i.e., the parameter's value when the game is in progress). The $\overline{\lambda}$ are parameters that the team most want to see. Based on these, we obtained the following:

\begin{equation}
Q(\lambda,\overline{\lambda})=\sum_IlogP(O,I|\lambda)P(O|\overline{\lambda}).
\end{equation}

In the case study, $P(O|\overline{\lambda})$ can be directly calculated from the initial original data. According to algorithm~\cite{abdel2014convolutional} in the HMM, we developed  $Q(\lambda,\overline{\lambda})$ by utilizing the known values, which is

\begin{equation}
\begin{split}
Q(\lambda,\overline{\lambda})=\sum_Ilog\pi_{i_1}P(O,I|\overline{\lambda})+
\sum_I(\sum_{t=1}^{\tau=1}loga_{j,j+1})P(O,I|\overline{\lambda})+\\\sum_I(\sum_{t=1}^\tau logb_{i_t}(o_t)P(O,I|\overline{\lambda}).
\end{split}
\end{equation}

Thus, three parameters have occurred in the formula. We deduce the values of these parameters as follows:

\begin{enumerate}
	\item[(1)] $\pi_i$:
	through the above calculations, we obtain 
	\begin{equation}
	\sum_Ilog\pi_{ij}P(O|\overline{\lambda})=\sum_{i=1}^Nlog\pi_iP(O,i_1=i|\overline{\lambda}).
	\end{equation}

	In the basketball matches,  someone is always chosen as an organizer in a round. Consequently, $\pi_i$ satisfies the following constraints:

	\begin{equation}
	\sum_{i=1}^N\pi_i=1.
	\end{equation}
	
	By using the Lagrange multiplier method, the Lagrange function is written as
	\begin{equation}
	\sum_{i=1}^Nlog\pi_iP(O,i_i=i|\overline{\lambda})+\gamma(\sum_{i=1}^N\pi_i-1).
	\end{equation}
	
	Take the partial derivative and make it 0.
	\begin{equation}
	\frac{\partial}{\partial \pi}\sum_{i=1}^Nlog\pi_iP(O,i_i=i|\overline{\lambda})+\gamma(\sum_{i=1}^N\pi_i-1)=0.
	\end{equation}
	
	We can obtain the following:
	\begin{equation}
	P(O,i_i=1|\overline{\lambda})+\gamma\pi_i=0.
	\end{equation}
	
	Then, we sum over $i$ and obtain $\gamma$
	\begin{equation}
	\gamma=-P(O|\overline{\lambda}).
	\end{equation}
	
	We can obtain $\pi_i$ as
	\begin{equation}
	\pi_i=\frac{P(O,i_i=i|\overline{\lambda})}{P(O|\overline{\lambda})}.
	\end{equation}
\end{enumerate}
\begin{enumerate}
	\item [(2)] $a_{ij}$: it is the same as  $\pi_i$. We sum over the probability that a player's pass in a game is 1, satisfying the constraint
	\begin{equation}
	\sum_{j=1}^Na_{ij}=1.
	\end{equation}
	
	By using the same method as (13), we obtain 
	\begin{equation}
	\sum_I(\sum_{t=1}^{\tau=1}loga_{j,j+1})P(O,I|\overline{\lambda})=
	\sum_{i=1}^N\sum_{j=1}^N\sum_{t=1}^{\tau-1}loga_{ij}P(O,i_t=i,i_t+1=j|\overline{\lambda}),
	\end{equation}

	\begin{equation}
	a_{ij}=\frac{\sum_{t=1}^{\tau-1}P(O,i_t=i,i_{t+1}=j|\overline{\lambda})}{P(O,\sum_{t=1}^{\tau-1}|\overline{\lambda})}.
	\end{equation}
\end{enumerate}

\begin{enumerate}
	\item [(3)] $b_j(k)$: the derivation process is the same as (2),
\end{enumerate}

\begin{equation}
\sum_{k=1}^Mb_{j}(k)=1,
\end{equation}

\begin{equation}
\sum_I(\sum_{t=1}^\tau logb_{i_t}(o_t)P(O,I|\overline{\lambda}))=\sum_{j=1}{N}\sum_{t=1}{\tau}logb_{j}(o_i)P(O,i_t=j|\overline{\lambda}),
\end{equation}

\begin{equation}
b_j(k)=\frac{\sum_{t=1}^{\tau}P(O,i_t=j|\overline{\lambda})I(o_t=v_k)}{\sum_{t=1}^{\tau}P(O,i_t=j|\overline{\lambda})}.
\end{equation}

	\section{Experiment and Results}\label{sec:CaseStudy}

We present the results of the computation of the methods for all games played by the Golden State Warriors and Cleveland Cavaliers in the 2016 season of the National Basketball Association (NBA). There are two reasons for choosing these two teams in this season. First, they were the most consistent teams in the league in recent years, and they did not change any aspects of their regular rotation during this season. Second, these two teams won all three championships during the past three years. According to common sense~\cite{mikolajec2013game}, we consider the Finals to be the environment that best shows the players' true characteristics and their value to their team. Furthermore, we can also analyze more players' data and reduce the volatility caused by changes in the lineup~\cite{blei2003latent}. More importantly, through rigorous mathematical derivations, we can compare the actual parameters for which we used real data from the finals, with the parameters trained in our system, to verify the accuracy of the training system that we obtained. Due to page limitations, we only analyze some games, which are randomly selected from all competitions, and we evaluate the exact values of two players (Stephen Curry and Andrew Bogut) in this paper. The results of other games and players are presented in the appendix .   

\subsection{Data}
The data for all games in the 2016 season of the NBA were downloaded from the SportVU website. SportVU is an optical tracking system installed by the NBA on all 30 courts to collect real-time data~\cite{siegle2013design}. We generate a sequence of points based on rounds, and we also present the data for each team as a table (Table~\ref{table1} shows 20 games for the Cleveland Cavaliers). In addition, we introduce artificial passing data (namely, Table~\ref{table2} shows the number of passes for the Cleveland Cavaliers), which include the Finals from the 2015 and 2016 seasons, to verify the parameters' effectiveness. 

We set up a Java web project to show all of our results. The passing network and graph were created and analyzed using Java and NetworkX~\cite{hagberg2008exploring}. The code implementation of the algorithm was created using Python.

\begin{table}[!ht]
		
		\centering
		\caption{ \bf The datasets of 20 games for the Cleveland Cavaliers.}
		\label{table1}
		\begin{tabular}{|c|c|c|c|c|c|c|c|}
			\hline 
			\bf	number & \bf date & \bf  round & \bf 0 pt ratio & \bf 1 pt ratio & \bf 2 pt ratio & \bf 3 pt ratio & \bf result \\ 
			\hline 
			1 & 2015/10/27 & 111&56.76
			& 9.01
			& 26.13
			& 8.11
			& F
			\\ 
			\hline 
			2 & 2015/10/28
			& 101
			& 48.51
			& 10.89
			& 27.72
			& 12.87
			& S
			\\ 
			\hline 
			3 & 2015/10/30
			& 107
			& 46.73
			& 16.82
			& 30.84
			& 5.61
			& S
			\\ 
			\hline 
			4 & 2015/11/2
			& 101
			& 43.56
			& 14.85
			& 33.66
			& 7.92
			& S
			\\ 
			\hline 
			5 & 2015/11/4
			& 115
			& 51.3
			& 20
			& 22.61
			& 6.09
			& S
			\\ 
			\hline 
			6 & 2015/11/6
			& 100
			& 47
			& 8
			& 35
			& 10
			& S
			\\ 
			\hline 
			7 & 2015/11/8
			& 108
			& 49.07
			& 15.74
			& 27.78
			& 7.41
			& S
			\\ 
			\hline 
			8 & 2015/11/10
			& 117
			& 40.17
			& 28.21
			& 22.22
			& 9.4
			& S
			\\ 
			\hline 
			9 & 2015/11/13
			& 109
			& 53.21
			& 16.51
			& 24.77
			& 5.5
			& S
			\\ 
			\hline 
			10 & 2015/11/14
			& 118
			& 54.24
			& 14.41
			& 19.49
			& 11.86
			& F
			\\ 
			\hline 
			11 & 2015/11/17
			& 100
			& 50
			& 12
			& 27
			& 11
			& F
			\\ 
			\hline 
			12 & 2015/11/19
			& 101
			& 36.63
			& 23.76
			& 28.71
			& 10.89
			& S
			\\ 
			\hline 
			13 & 2015/11/21
			& 104
			& 45.19
			& 15.38
			& 28.85
			& 10.58
			& S
			\\ 
			\hline 
			14 & 2015/11/23
			& 103
			& 45.63
			& 12.62
			& 24.27
			& 17.48
			& S
			\\ 
			\hline 
			15 & 2015/11/25
			& 98
			& 50
			& 13.27
			& 22.45
			& 14.29
			& F
			\\ 
			\hline 
			16 & 2015/11/27
			& 106
			& 50
			& 17.92
			& 24.53
			& 7.55
			& S
			\\ 
			\hline
			17 & 2015/11/28
			& 104
			& 55.77
			& 10.58
			& 25
			& 8.65
			& S
			\\ 
			\hline 
			18 & 2015/12/1
			& 106
			& 54.72
			& 18.87
			& 17.92
			& 8.49
			& F
			\\ 
			\hline 
			19 & 2015/12/4
			& 113
			& 47.79
			& 17.7
			& 25.66
			& 8.85
			& F
			\\ 
			\hline 
			20 & 2015/12/5
			& 109
			& 56.88
			& 14.68
			& 22.94
			& 5.5
			& F
			\\ 
			\hline 
			
		\end{tabular}
		\begin{flushleft} The table is processed from the data of a part of our appendix.
		\end{flushleft}

\end{table}

\begin{table}[!ht]

		\centering
		\caption{ \bf The number of passes in the finals of the Cleveland Cavaliers.}
		\label{table2}
		\begin{tabular}{|c|c|c|c|c|c|c|c|c|c|c|c|c|c|c|c|c|c|c|}
			\hline
			& \bf A & \bf  B & \bf  C & \bf  D & \bf  E &  \bf F & \bf  G & \bf  H & \bf  I & \bf  G & \bf  K & \bf  L & \bf  M\\
			\hline
			\bf 	A & 95 & 36 & 20 & 17 & 112 & 13 & 7 & 3 & 2 & 2 & 2 & 0 & 1\\
			\hline
			\bf 	B & 20 & 78 & 4 & 9 & 54 & 15 & 2 & 11 & 1 & 2 & 0 & 1 & 1\\
			\hline
			\bf 	C & 10 & 0 & 65 & 2 & 17 & 5 & 1 & 2 & 0 & 1 & 0 & 0 & 0\\
			\hline
			\bf 	D & 13 & 5 & 1 & 95 & 35 & 11 & 1 & 18 & 0 & 3 & 0 & 0 & 1\\
			\hline
			\bf 	E & 79 & 73 & 45 & 52 & 411 & 99 & 24 & 84 & 6 & 35 & 2 & 0 & 21\\
			\hline
			\bf 	F & 20 & 8 & 7 & 16 & 55 & 121 & 3 & 12 & 1 & 7 & 0 & 0 & 6\\
			\hline
			\bf 	G & 6 & 1 & 3 & 0 & 9 & 8 & 18 & 4 & 0 & 0 & 0 & 0 & 0\\
			\hline
			\bf 	H & 4 & 13 & 6 & 19 & 108 & 32 & 5 & 161 & 0 & 29 & 0 & 1 & 12\\
			\hline
			\bf 	I & 2 & 0 & 1 & 0 & 3 & 2 & 6 & 0 & 6 & 0 & 0 & 0 & 0\\
			\hline
			\bf 	J & 2 & 0 & 0 & 2 & 22 & 5 & 0 & 11 & 0 & 48 & 0 & 0 & 0\\
			\hline
			\bf 	K & 1 & 0 & 0 & 0 & 0 & 0 & 0 & 0 & 0 & 0 & 3 & 0 & 0\\
			\hline
			\bf 	L & 0 & 0 & 0 & 0 & 0 & 0 & 0 & 0 & 0 & 0 & 0 & 1 & 1\\
			\hline
			\bf 	M & 2 & 1 & 1 & 2 & 10 & 7 & 0 & 4 & 0 & 0 & 0 & 0 & 32\\
			
			\hline
		\end{tabular}
		\begin{flushleft} The table is processed from the data of a part of our appendix. The letters (A $\sim$ M) represent the players of the Cleveland Cavaliers.
		\end{flushleft}

\end{table}

\subsection{CaseStudy}

In this section, through a verification experiment, we will prove that the proposed method, which combines networks with a hidden Markov model, is a feasible approach. The experimental results are more accurate and stable for evaluating the exact values of basketball players.

In this experiment, we have 13 players recorded during the season for the Golden State Warriors, for whom are incorporated the semantics of the basketball games. Now, we indicate them with letters (e.g., A represents Stephen Curry). Each round is given a set of 4 results (0 points, 1 point, 2 points, and 3 points). Thus, we set the known parameters in the model as follows: $\mathcal{M} = \left\lbrace A, B, C ... L, M\right\rbrace$. $\mathcal{N} = \left\lbrace 0,1,2,3 \right\rbrace$. The observation sequence $O_{i}$ of our selected games is inputted into the algorithm. We can obtain the parameters $A$ (matrix of state transition probability), $B$ (matrix of observation probability), and $pi$ (initial probability distribution). According to the above analysis, the parameters that were not directly obtained from the original large dataset can be derived from the algorithm in our system. By using these parameters, we provide a novel behavior analysis and individual evaluation method in the basketball game. Now, we combine the method with the data from the basketball competition and introduce the experiment in detail from two perspectives.

\begin{Method}
	The method for verifying the accuracy of the parameters trained in VPHMM
\end{Method}

It is well known that in hidden Markov models, if the parameter of a state transition matrix probability is the same, the other two parameters are the same. This is determined by its derivation formula.~\cite{lecun2015deep}. In other words, if we prove that the parameters $A$ (matrix of state transition probability), which is a training parameter, is the same as its defined parameter, which is generated by the actual numbers, then we can verify the effectiveness of the parameters trained by the system. To obtain the parameters in the definitions, we obtained the passing data for the NBA Finals in 2015 and 2016 through artificial collection (such as Table~\ref{table1}). In this way, we can calculate the metric of state transition probability from its definition and compare the results with those trained by our system. Moreover, we have a graphical processing~\cite{kumar2016mega7} for showing the data intuitively. 

\begin{figure}[!ht] 
	\centering 
	
	{ 
		\label{fig:subfig_a25} 
		\includegraphics[width = 1\textwidth]{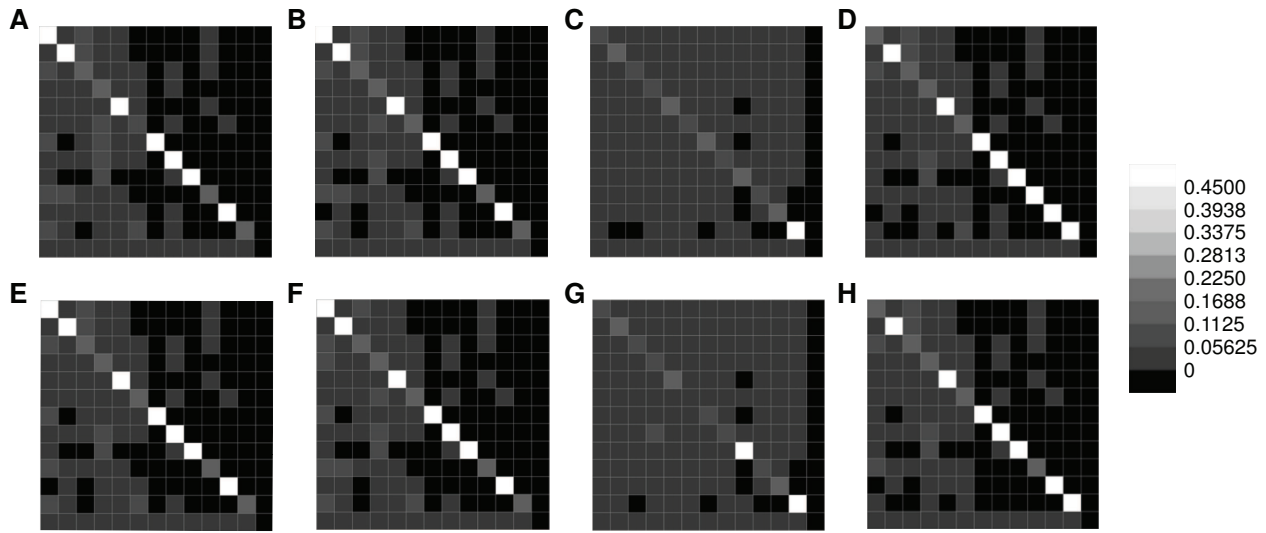} 
	} 
	\caption{{\bf There are 8 figures, which represent 13 players' relationships in four games. The scale on the right-hand side of each figure displays the probability that player X will interact with player Y.} There are four groups (a and e, b and f, c and g, and d and h), which are a comparison of the matrix of transition probability. Figures (a $\sim$ d) are authentic data, and (e $\sim$ h) are trained by the algorithm.}
	\label{fig3}
\end{figure}

Fig~\ref{fig3} illustrates the probability that players X and Y interact (i.e., passing the ball to each other) in four games. The more interactions that occurred among players, the lighter is the color displayed in the grid, as displayed in the reference scale on the right-hand side of each grid. There are eight pictures here, each of which is composed of 13 $\times$ 13 grayscale lattices. They are divided into 4 groups. Both figures in one group are printed to describe the same game, whose parameters are calculated in two different ways. The top row is the result of the real observation, and the bottom row is the result of our algorithm. 
By comparing and analyzing the data from the four groups (see Fig~\ref{fig3}), it can be observed that deviation of the parameters is in the acceptable range~\cite{cai2019personalized}. In our experiment, we carefully compared all the game parameters of different teams in the 2016 and 2017 NBA Finals. Thus, we can consider that parameters trained in our system are applicable and the method of training parameters is feasible.

\begin{Method}
	A novel criterion to evaluate the exact values of players.
\end{Method}
Duch et al.~\cite{duch2010quantifying} proposed that evaluating any network of a team requires determining the relative value of its individual members. In the same way, the value of a player also depends on the network of his team. The network here is directly determined by the interaction between players on the basketball court. To solve this problem in a comprehensive way, we propose a novel criterion about basketball by quantifying the involvement of players in the network and calculating the probability distribution of their points and the probability that the player was chosen as an organizer with successful versus unsuccessful results. For our analyses, we used the parameters that were trained from our algorithm to compare the different players. We provide the methodology to evaluate the exact values of players and illustrate its concrete steps with two players (i.e., Stephen Curry and Andrew Bogut) of the Golden State Warriors.

Through the training system, we obtain three parameters (matrix of transition probability $A= \{ A_{ij}$\}, matrix of observation probability $B$ = \{$b_{ij} $\}, and initial probability distribution $\pi$ = \{$\pi_{i}$\}) of the Golden States Warriors in the 2016 season. In our Methodology, we have defined the parameters $A$, $B$, and $\pi$. Now, we visualize each of the parameters to obtain the values of the players in the comparison. From the data visualization, the matrix of transition probability is indicated as a weighted graph (Fig~\ref{fig1}), the matrix of observation probability is shown as a pie chart (Fig 4), and the initial state probability distribution is expressed as a bar chart (Fig 5). In the case study, we randomly choose two games of the Golden State Warriors to evaluate Stephen Curry's and Andrew Bogut's exact values to the team~\cite{kearse2012geneious}.

For the purpose of evaluating the exact values of Stephen Curry and Andrew Bogut of the Golden State Warriors, we generate the network representation for every game involving the Golden State Warriors. Figure 1(a) is one of the games chosen randomly among the Warriors' wins. Meanwhile, Figure 1(b) is one of the games chosen randomly among the Warriors' losses. In this paper, we take these two games, combined with the three features of the parameters, as the examples suggest. Before the units of each parameter, we first compare two weighted graphs. There is a clear difference in the density of passes between the successful games and the lost game. This provides an important reference for our next analysis. The evaluation method is instantiated in the parameters as follows:
\begin{enumerate}
	\item 
	By using data from the matrix of transition probability, we creatively construct the individual weight graphs, which show each player's linkages from the player's point of view. Figure 6 shows the networks of Stephen Curry; we can observe that the networks are very different between the successful game and the lost game. We analyze the results of the game from two perspectives.
	\begin{itemize}
		\item 
		According to the comparison between the two games, we can observe that when the game is won, the probability of Curry's passed balls and received balls are more balanced with all his teammates. In the lost games, he was more inclined to interact with several core players. In other words, his interaction with his teammates is very important to the game. 
		\item 
		In addition, whether the game results are wins or losses, Stephen Curry can contact almost all of the players. Figure 7 presents the network of another Golden State Warriors' player named Andrew Bogut. Clearly, we can intuitively see that his passed arrows and received arrows in the two games are both less than those of Stephen Curry. We can observe that this player prefers to obtain the received balls, which means that this player is more dependent on assists from his teammates than on creating an opportunity for himself~\cite{blanchard2009cohesiveness}.
	\end{itemize}
	\item From the data visualizations, the matrix of observation probability was presented as a pie chart. This pattern showed the significance of Stephen Curry to the team.
	Comparing the two charts (i.e., Figure 4a and Figure 4c), we can intuitively see a great difference in the performance of Stephen Curry between the two games. When he did not play at his proper level (e.g., Fig 4c), the probability of losing the game was larger. It was also observed that Andrew Bogut's performance in the two games was almost identical (see Fig 4b and Fig 4d).
	\item According to the initial probability distribution, we transformed the probability distribution into a bar chart. The chart shows the probability of players in the game being the organizer of each round. By analyzing Fig 5, we can analyze the probability changes of all players (e.g., Stephen Curry and Andrew Bogut) during the different games. In Fig 5a and Fig 5b, the X-axis represents each player, and the Y-axis indicates the probability that a player is chosen as the organizer. 
	By analyzing the two histograms, we can obtain a clear conclusion. For the team, Stephen Curry is the organizer who the team relies on in every game. Moreover, through the comparison of probability, a higher proportion of passes involving Stephen Curry is propitious to the game's success. By contrast, the proportions of the other players (e.g., Andrew Bogut) are relatively small and have little influence. 
\end{enumerate}

We drew some interesting conclusions by comparing the two players in the two different games. All players from different teams can be added to the comparison in this way. From the results, it was possible to visually identify the differential influences of the performances of various players on the result of the game. The quantification and visualization of the value of a player to his team can provide coaches and managers with useful knowledge for deciding on the roster of their teams.
	\section{Conclusion}\label{sec:Conclusion}
In this paper, we presented an approach to evaluate the exact values of basketball players to their team using SportVU data collected from the NBA. For this purpose, we first defined and trained three parameters, depending on the passing network, the probabilities for a person being chosen as an organizer, and the distribution of the player's scores. To verify the effectiveness of the parameters trained in the system, we used a mathematical derivation for the formula to estimate the parameters and calculate the deviation between the output results of the algorithm and the real data. We accomplished this by using an algorithm called the variant parameter hidden Markov model (VPHMM), which was used to analysis the players' networks.
To obtain better evaluation performance, we respectively visualized the three parameters as a weighted graph, a pie chart, and a bar graph. The overall goal of this work was to create an assistive tool for the manager of the teams, coaches, and players to use when planning to improve the team and play against an upcoming opponent. The experiment showed that our method of evaluating basketball players' values in teams is superior to previous strategies. We believe this method can play a role in other group sports.

	\section*{Acknowledgments}
	The research work described herein was funded by the Science and Technology Planning Project of Guangdong Province (No. 2016B010124012, 2016B090920095).

	\section*{References}
	\bibliography{references}
	
\end{document}